# Planning for Directory Services in Public Key Infrastructures

Vangelis Karatsiolis
Technische Universität Darmstadt,
Darmstadt Centre for IT Security
Hochschulstr. 10, 64289 Darmstadt, Germany
karatsio@cdc.informatik.tu–darmstadt.de

Marcus Lippert, Alexander Wiesmaier
Technische Universität Darmstadt,
Department of Computer Science,
Hochschulstr. 10, 64289 Darmstadt, Germany
mal,wiesmaie@cdc.informatik.tu–darmstadt.de

**Abstract:** In this paper we provide a guide for public key infrastructure designers and administrators when planning for directory services. We concentrate on the LDAP directories and how they can be used to successfully publish PKI information. We analyse their available mechanisms and propose a best practice guide for use in PKI. We then take a look into the German Signature Act and Ordinance and discuss their part as far as directories concerning. Finally, we translate those to the LDAP directories practices.

## 1 Introduction

Public key infrastructures (PKIs) play a significant role in securing today's communication. Entities use certificates that enable security mechanisms like confidentiality, integrity, non–repudiation and authenticity. There are two specifications for such certificates, namely the X.509 [IT97] and the PKIX [HPFS02]. The first specification gave its name to these certificates which are called X.509 certificates. The goal of X.509 was to enable authentication mechanisms for directories [IT97, Sec. 1]. But X.509 certificates are used today in various cases of internet security, for example in S/MIME to secure electronic mail [Ram04].

X.509 certificates are issued by a certification authority (CA) and they are public documents. This fact suggests that they should be published in a public directory. The most commonly used directories for this purpose are based on LDAP [HM02]. LDAP directories are used as the central place in a PKI, where certificates and associated revocation information, in the form of certificate revocation lists (CRLs), are stored and can be downloaded from various clients.

This paper is organised as follows: In Section 2, we take a look at the use of LDAP in PKI. We discuss the certificate and CRL publishing, the security features of LDAP and different issues related to the planning of LDAP to support PKI. In Section 3, we investigate the relationship of the German Signature Act and Ordinance to LDAP directories. We then discuss the LDAP related practices in this electronic signature law context. Lastly, in Section 4, we conclude the paper.

## 2 LDAP and PKI

LDAP directories are used to hold certificates and CRLs in order to provide dissemination of PKI information. There are other mechanisms to enable this like HTTP or FTP (for a discussion see [HP01, Chap. 9] and [AL99, Chap. 11]). All these mechanisms are scalable and standardised, but LDAP is the most commonly deployed solution. Many organisations already operate such directories. A report for using Microsoft's Active Directory, which supports LDAP, along with PKI in the corporate environment for 300.000 users is found in [GSB+04]. In addition, typical clients (like email clients) already have LDAP interfaces to retrieve CRLs and certificates. Nevertheless, the way that LDAP behaves in some cases, complicates its support to PKI. Chadwick [Cha03] makes an in–depth study on this behaviour. Gutmann [Gut00] suggests that a relational database is a better solution than LDAP as far as storing of certificates and CRLs concerning. In this paper we concentrate on the use of LDAP, since this is a solution already employed by many PKI practitioners. One of them is RegTP,[1] the authority responsible for the German root CA, according to §3 of the German Signature Act[2] [Leg01a]. RegTP uses an LDAP directory to provide public availability for qualified certificates as §2.12.b SigG requires.

### 2.1 Storing certificates

#### 2.1.1 User certificates

Information on an LDAP directory is hierarchically organised. In Figure 1 we can see the entry of the user `CN=Alice, O=Org, C=DE`. The values between the commas are called relative distinguished names (RDN) and the whole value a distinguished name (DN). Every entry has a relative distinguished name. If someone moves from this entry to the root, and adds the relative distinguished name of every entry he meets, the distinguished name for this entry is built. Relative distinguished names are usually in the form `attribute=value` but they can also be multi–valued and presented like `attribute1=value1+attribute2=value2`. This representation gives the ability to create unique RDNs among siblings in the directory. For example, in the case of certificates the issuer name and the serial number form together a unique representa-tion [HPFS02, Sec. 4.1.2.2]. This representation could be used on the LDAP to create these unique entries in the form of a multi–valued RDN. Such a representation is work in progress at the IETF [IET].

---

[1] It stands for Regulierungsbehörde für Telekommunikation und Post.
[2] We abbreviate the German Signature Act as SigG from the German term which is Signaturgesetz.

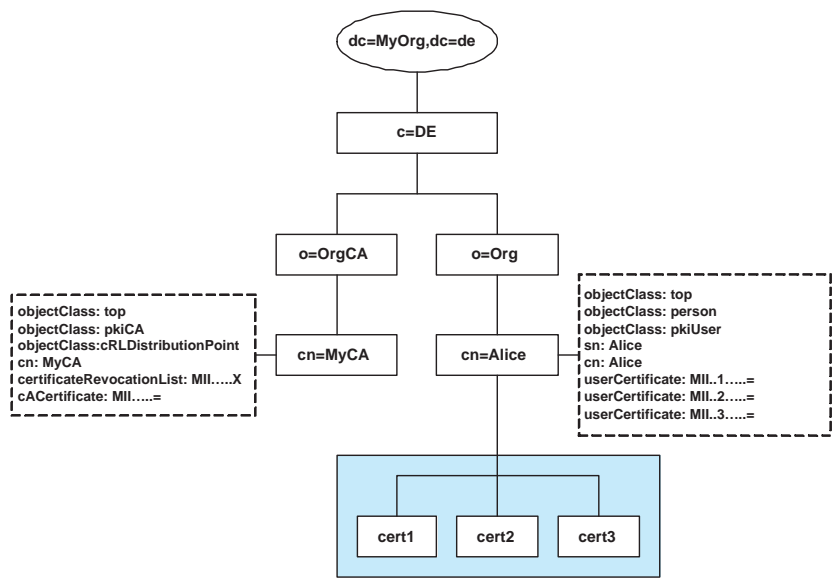

Figure 1: A user and a CA entry in the directory. Beneath the user entry, the idea of having separate entries for each certificate is presented.

Every entry on the LDAP has attributes and every attribute has one value, or more in case of not single–valued attributes (see [WCHK97] for details). All entries have the attribute called *objectClass*. This attribute describes the kind of the entry on the LDAP. Based on the objectClass attribute values, the entry can have different properties. Different properties can be applied to the entry by extending the values of the object class. In Figure 1 the attributes for an end–user and a CA entry are shown. Usually, but not necessarily, the subject distinguished name contained in the entry's certificate matches the distinguished name of the entry in the directory. In the end–user entry the *person* object class as well as the *pkiUser* object class are present. The first one gives the entry the ability to hold attributes like the surname or the telephone number. The second one enables the entry to hold the *userCertificate* attribute which represents the X.509 certificate for this entry. The userCertificate attribute holds the certificate in its binary form in the directory.

The PKI developer and administrator should make a careful decision on the schema (see [WCHK97, Sec. 3] for the definition) they want to use for publishing certificates. The solution should not make any assumptions on the current status of the directory as well as it should not be planned to support only PKI. LDAP directories can be used to hold information not related to PKI. One choice is to use the *strongAuthenticationUser* object class [Wah97]. The drawback with this approach is that this object class requires that a certificate is present on the entry and therefore it can not be used in cases where the entry does not possess any certificates (for example, the certificate is removed from the directory after its revocation). In addition, this object class refers to the X.509 strong authentication mechanism and it should not be used just as a certificate container. The solution is to use the *pkiUser* object class [BHR99]. This attribute allows but does not require that the userCertificate attribute is present. Therefore the previous drawback does not apply in this case. Lastly, the *inetOrgPerson* object class allows the certificate to be published on the entry too. This object class may be used when the directory is not only used to support PKI but other organisational procedures and data.

### 2.1.2 CA certificates

The object classes used to enable publishing of CA certificates differ to the ones for end users. There are two choices. The first one is the *certificationAuthority* [Wah97] and the second one is the *pkiCA* [BHR99]. The certificationAuthority mandates that a CRL, the CA certificate as well as an authority revocation list[3] (ARL) [IT97] must be published on the directory. But very few CAs issue ARLs or the CA represented on this node may decide not to issue CRLs itself. Therefore, this object class is not the best choice and the more flexible pkiCA should be used. This object class allows CRLs, ARLs and CA certificates to be published. The CA certificate is represented from the *caCertificate* attribute, the X.509 CA certificate in its binary encoded form. Lastly, both object classes allow the *crossCertificatePair* attribute. This attribute represents the forward and reverse (or issuedToThisCA and issuedByThisCA) cross–certificate pair for cross certification purposes.

### 2.1.3 Certificate search

Certificates in LDAP are stored as binary objects at the entry of their corresponding entity. Therefore locating the correct certificate forces the client to know this entry (probably from an email search or the subject distinguished name), download the certificate, then parse it and lastly examine whether this is the one it searches for or not. This process however, delegates the certificate search to a some kind of side–information search, in fact some information related to the owner of the certificate and not the certificate itself.

A better approach is to publish information related to the certificate, in order to enable searching for this data. There is a work in progress in the IETF [IET] that specifies such meta–data that are published together with the certificate. This work is easy to implement, since the attributes can be extracted from the certificate and published along with it. This simplifies the search for certificates with special attributes, since now on the node the properties of the certificates (like serial number, subject DN etc.) are held as attributes of the node. This solution does not incorporate cross certificates from one CA to another (cross certification) and future work should address this problem. Another work in progress in IETF [CL] specifies special matching rules on the directory to locate the correct certificates. This work is more extensive than the first one but more difficult to implement. This implementation must take place at the side of the LDAP vendors and clients.

---

[3] A list associated with revocation information only for CA certificates.

The first work proposes also that every certificate will have its own entry subordinate to the user entry (for a simple and not detailed draw of this idea, see Figure 1). This can help solving the following problem. When a user owns more than one certificate, this is arranged in LDAP with a multi–valued attribute, namely the userCertificate. A problem with this approach arises if someone wants to locate only one specific certificate for this user. Then, one must download all certificates and find out the correct one by searching for the correct certificate locally. But if every certificate has its own entry then this problem does not apply anymore. A more generic solution to this problem is a new standard from IETF [CM04] which defines a method, that enables matching of the values (and not of the attributes which is common in LDAP) with a special filter associated only with the values. Current practice should orient on the subentries solution since most LDAP servers and clients do not support the special matching rules and returning of matched values. Long term planning however, should incorporate the most generic solutions which are on the other hand vendor and client dependent. Similar solutions can be applied to CA certificates.

All solutions discussed here, should be used carefully with regard to reliability of the information contained on the LDAP server. This information is not signed information and it could have been manipulated from an attacker. Every client must verify the signature on the certificate and CRL to ensure the security of the data. Nevertheless, LDAP has a number of security features which can be turned on to guarantee that no unauthorised changes occur in the directory. It can also then meet certain requirements of the Signature Act and Signature Ordinance [Leg01b].

## 2.2 Security

Various security mechanisms can be applied to LDAP directories. They address different directions and therefore cover many security aspects of an on–line electronic system. These attractive security features make LDAP directories a perfect candidate for the directory services in the SigG context. First of all they allow different means of authentication. One is the simple authentication with password. Another possibility is to combine it with SASL [Mye97] for a digest based authentication. For a description see [WAHM00]. Lastly, typical TLS client authentication is also possible.

TLS [DA99] can also be used for securing the network traffic to the LDAP server. Of course, the server authentication is a necessary step in this protocol, and therefore the LDAP server must authenticate itself to the clients. TLS can be used to avoid that the passwords will travel in clear in the simple password authentication. For more on this scheme see [HMW00].

Apart from the authentication and security mechanism, access control mechanisms are applied to LDAP. The directory administrator can set access control lists (ACLs), in order to allow certain actions to definite individuals and special parts in the directory. This scheme is very strong and allows granular controls concerning persons, actions, and data in the directory. An application of the above could be that a revocation authority is able to publish only CRLs on the directory, while the certification authority (CA) only certificates. In [Cha00] and [Gre02, Chap. 5] more on the security of LDAP directories can be found.

There are several cases, in which these security mechanisms should be applied. For example, in order to avoid that an unauthorised user replaces a revocation list with an older one, or just removes a certificate. Although certificates and CRLs are signed information and therefore their correctness can be proved, the integrity of the data on the directory is very important in order to prevent such a misuse. In addition, using the TLS mechanism of LDAP, the server can authenticate itself. Thus, the publishing client can check whether it publishes the information on the correct server. This ensures that the certificates and revocation information are published in the correct place, where the clients are expecting this information to be published.

### 2.3 Storing CRLs

CRLs are stored in a special LDAP attribute called *certificateRevocationList* defined in [BHR99]. There are three object classes that allow this attribute, namely the *certificationAuthority* from [Wah97], the *pkiCA* and the *cRLDistributionPoint* defined in [BHR99]. The first one can not be used when indirect CRLs [HPFS02, Sec. 5] are utilised, since the entry on the directory is not a CA at all and no CA certificate should be present. The other two object classes allow the certificateRevocationList attribute without mandating this or other attributes. In addition the cRLDistributionPoint object class mandates the *commonName* attribute which gives the ability to build an entry on the LDAP with the CN node. The best choice is a hybrid solution where both object classes are present (not necessarily on the same entry). The pkiCA is used then to publish the CA and cross certificates and the cRLDistributionPoint to publish any revocation information.

In case of indirect CRLs the cRLDistributionPoint object class is the only choice. The CRL issuer in this case is not the issuer of the certificate and the keyUsage on its certificate is for CRL signing. This CRL issuer does not possess any CA certificate and the pkiCA object class as a choice would be just wrong, since it denotes a certification authority. In the case of the German root authority the CRLs issued are indirect ones. This is an implication of the special validity model of the SigG. For more on this validity model see [Bau98].

In an X.509 certificate there is the possibility to state inside the certificate where to find revocation information about it. This is arranged in a special extension called *cRLDistributionPoints* [HPFS02, Sec. 4.2.1.14]. In this extension the place where clients can locate the CRL related to this certificate is provided. This can be an LDAP URL and/or a HTTP URL, among others, which implies also the mechanism used to obtain the CRL. Use of this extension is recommended since it simplifies the revocation information management. The LDAP URL for the entry on the directory in Figure 1, according to the format defined in [HS97], is:

```
ldap://192.168.0.1:389/CN=MyCA,O=OrgCA,C=DE,DC=MyOrg,DC=DE?
certificateRevocationList?base?objectClass=cRLDistributionPoint
```

Delta CRLs can be published in the LDAP directory, too. The corresponding attribute is the *deltaRevocationList* and an additional object class that can hold this attribute is the *deltaCRL* [BHR99]. The cRLDistributionPoint object class can also hold delta CRLs. However, the way this information is arranged in the LDAP directory does not utilise the idea of delta CRLs, since these become meaningful only when the base CRL is also located in the directory. Every time a delta CRL is issued a full CRL must be issued, too. This is done in order for clients that can not handle delta CRLs to be able to have the freshest revocation information. This was a mandatory step in the deprecated PKIX specification for X.509 certificates [HFPS99]. This suggests that the newer CRL replaces the base CRL (it is typical that the newer CRL just overwrites the older one in the directory). This is not a problem if the clients already have the base CRL in their cache. But new clients (or even clients which were off–line when the base CRL was placed in the directory) have to work with complete CRLs until they have the chance to locate a base CRL in the directory. To overcome this, the newer CRL must be just added to the directory and not replace the older one. In this case however, the clients must download all CRLs in order to find the base CRL. In the newer PKIX specification [HPFS02], when a delta CRL is published, it is mandatory that the corresponding base CRL must be found in the directory. An LDAP mechanism to distinguish between the different CRLs is needed. The new scheme for CRLs proposed in IETF [IET], will support this and therefore this problem can be solved. In this scheme every CRL is stored in its own entry, in a similar way to the certificate proposal.

### 2.4 Naming plan

In order to arrange the information effectively in the hierarchical manner of an LDAP directory there is a proposal in RFC 2377 [GHSW98]. This RFC proposes the use of the domain component attribute for the root of the directory. Domain names are unique and hierarchical. Therefore they can be used in order to provide unique names in the LDAP model. In [KWG$^+$98] the appropriate object classes to enforce this proposal can be found. In the Active Directory of Microsoft, the use of domain names as the root of the directory is mandatory [Gre02, Chap. 6]. Active Directory must be installed and configured properly in order to use the integrated CA shipped with Windows 2000 Server.

### 2.5 Other uses of LDAP

LDAP directories can be used of course for other purposes too. As already mentioned, many organisations already operate LDAP directories to administrate and manage their information. A common LDAP use is for central authentication purposes in UNIX systems. In [KLW04] it is shown that LDAP directories can be used for providing proof–of–possession for encryption keys as well as delivery of software personal security environments (PSE). In those schemes there are two applications which come along. One is the activation of certificates. With the term activation we denote the action to make the cer-tificates public available. The other application is the private key on demand, in which the user can locate and download his PSE whenever he wants to use it. Guida et al. [GSB+04] have used the Active Directory for the registration purposes inside the PKI, in order to extract identity information for the issuing of certificates.

## 3 Directories and the German Signature Act

Qualified certificates must be verifiable and optionally available. With verifiable it is implied that a mechanism to determine whether this certificate exists as well to obtain information about its status is present. With available it is implied that the certificate itself is located in a position where everybody can find and download it. Every Certification Service Provider (CSP) therefore, must provide a directory service where certificates and associated revocation information is located. One complete technical solution to this consists of an OCSP server [MAM+99] to keep the certificates verifiable and an LDAP server to have the certificates available.[4]

OCSP servers give signed answers to a client's query about the status of the certificate. There are three possible answers. The first one is good, meaning that the certificate is not revoked. The second one is revoked, meaning that the certificate is revoked and the third unknown, meaning that the certificate does not exist for this OCSP server (it is unknown to it). Therefore OCSP can be used to give information about the status of a certificate. The special extension *CertHash* has been specified by the ISIS–MTT specification [TT04], in order to include also an existence evidence of the queried certificate. This extension holds the hashed value of the certificate the clients want to verify. With this OCSP configuration the OCSP meets the verifiable requirement of the law. According to §15.2.2.b SigV, the verification of a qualified signature requires also a proof of existence of the qualified certificate, as well as information about its status, at the signature creation time. Further we show how LDAP can be used and be appropriate in the SigG context to provide availability of qualified certificates. But more mechanisms can also be employed in parallel. A candidate for example is the HTTP store as described in [Gut00]. There is a work in progress at IETF [IET] that specifies this scheme. The above discussion demonstrates that any technical requirements are outside the scope of SigG. This is because the standards and techniques used to satisfy the law are subject to changes.

According to §3.1 SigV, no certificate must be published before its user is identified and accepts his qualified certificate. This means that compliant certification service providers must wait for an activation of the certificate from its user. A typical certification process in the corporate or institution environment usually automates the publishing of certificates. This automation however, must not be performed in the SigG context. Complying implementations should enable an activation mechanism for the publishing of certificates. Only after the user has accepted his certificate all signatures that he has calculated, even those before the acceptance, are considered in effect. When a certificate is activated, it should be verifiable and if its user requires it, also available. This implies that a certificate upon ac-tivation must become unconditionally known to the OCSP but conditionally be published on the LDAP server. Conforming CSPs must distinguish between certificates that must be kept public and those which must not and have the proper mechanism to resolve this.

---
[4]This is the current practice in RegTP.

The above is also discussed in §5.2 SigV. The key owner must explicitly accept his secure signature creation entity and confirm this to the CSP. Then it is the CSP's task to publish the certificate in order to keep the certificates publicly available and verifiable by the public (§2.12.b SigG). After the confirmation from the user, the CSP is obliged to keep the certificate on the directory for at least five years after its expiration year (§4 SigV). In the case of accredited CSPs this time window is thirty years after the certificate expiration.

Compliant publishing components should therefore call only add functions to the directory but never delete functions. In LDAP this is translated to calling add requests and never call delete or modify requests (for the definition see [WHK97]). This gives the ability to the developer to ensure correct functioning and compliance to the directives. If any remove action must be performed (for example after the thirty years), then this should be done manually and with the presence of a special monitoring mechanism. This mechanism will inform the administrator of his actions. For example, this mechanism will first fetch the certificate from the LDAP before the delete request, demonstrate this to the administrator, who now must check whether this is the correct certificate or not, and delete it after the administrator's acknowledgement.

Moreover, a special mechanism should exist that notifies in case of a failed operation. If an intended LDAP operation is not successful, the system must notify the administrator for the erroneous behaviour. The problem must be explicitly stated, in order to give the administrator a clear view of which step was not performed. Then, an out–of–band mechanism should be used to perform the unsuccessful step. It is clear that this out–of–band mechanism must have the same security properties as the regular one. Apart from this, a regular backup of the data on the directory must be performed by creating backup files of the whole directory. This can be done with the help of the LDAP data interchange format (LDIF) [Goo00]. LDIF is a text format that describes the data on an LDAP directory. The importance of the data persistence in the directory is depicted also in §15.3 SigV. The certificate presence as well as revocation information related to it must have a continuous state in the directory. In addition, in the case of an accidental error or serious system crash, the status of the directory can be retrieved as it was before this event.

According to §7.2 SigV, if a certificate is revoked this information must be published in the directory. In the LDAP case this is done by storing the CRL. As in the case of certificates, it must be ensured that the CRL publication was successful and if not, special countermeasures must be enforced. In the CRL publishing however, the activation mechanism is not needed. In this case it is very important that as soon as the CRL is produced from the CA,[5] this CRL is also published in the directory without any delay. Another difference is that in the CRL case a modify request is usually called, in order to replace the existent CRL in the directory. There is no need to keep all CRLs in the directory since the newest CRL contains revocation information of the previous CRLs within the same scope. If delta CRLs are used however, at least the base CRL should be found in the directory and not be deleted.

---

[5]Or a revocation authority since the CRLs used in the SigG context are indirect CRLs.

The access control mechanism of the LDAP is useful in this case, in order to distinguish between a CRL and a certificate publishing. The ACLs in the directory can be configured so that a special certificate publishing user is only allowed to add certificates (namely the userCertificate attribute) and in a special path in the directory. Another user, the CRL user, can publish only CRLs (namely the certificateRevocationList attribute). The other security mechanisms must be used too. The TLS connection guarantees the identity of the LDAP server and the client authentication ensures the identity of the user publishing data on the LDAP. Then, no unauthorised actions on the LDAP can be performed (like removing certificates or replacing CRLs) and moreover, publication takes place only on the correct directory.

## 4 Conclusion - Future Work

We analysed the LDAP directories and their use in PKI. LDAP is a reliable technical solution to address the PKI information dissemination problems. We took a look especially in the SigG environment. We discussed a planning for LDAP, in order to meet the special SigG requirements associated with the public availability of certificates and revocation information. A lot of work is in progress at the IETF trying to solve problems related to LDAP and PKI. A good planning for LDAP, promises to provide a successful support to PKI as well as to other organisational procedures, particularly when the new proposals become a standard and incorporated in the current solutions.

We plan to design and implement an LDAP PKI client for management of certificates, CRLs and software PSEs. A new standard is specified in [MSNP04] which defines a synchronisation method of an LDAP client with the LDAP directory. We will investigate how this technique can be used to automatically download new CRLs and how certificates for users which are already installed locally can be updated.

## Acknowledgements

The first author thanks Georgios Raptis for his help and guidance through the German Signature Act, and Markus Ruppert for discussions on management problems in the directory.

## References

[AL99]   C. Adams and S. Lloyd. *Understanding Public-Key Infrastructure*. New Riders Publishing, 1999.

[Bau98]  M. Baum. Gültigkeitsmodell des SigG. *Datenschutz und Datensicherheit*, 23(4):199–205, 1998.

[BHR99]   S. Boeyen, T. Howes, and P. Richard. Internet X.509 Public Key Infrastructure LDAPv2 Schema. *IETF Request For Comments*, 2587, June 1999.

[Cha00]   D. W. Chadwick. Secure Directories. In *Proceedings of the NATO Advanced Networking Workshop on Advanced Security Technologies in Networking*, June 2000.

[Cha03]   D. Chadwick. Deficiencies in LDAP when used to support PKI. *Communications of the ACM*, 46(3):99–104, 2003.

[CL]   D. W. Chadwick and S. Legg. Internet X.509 Public Key Infrastructure LDAP Schema and Syntaxes for PKIs. http://sec.isi.salford.ac.uk/download/pkix-ldap-pki-schema-00.txt.

[CM04]   D. Chadwick and S. Mullan. Returning Matched Values with the Lightweight Directory Access Protocol version 3 (LDAPv3). *IETF Request For Comments*, 3876, September 2004.

[DA99]   T. Dierks and C. Allen. The TLS Protocol Version 1.0. *IETF Request For Comments*, 2246, January 1999.

[GHSW98]   A. Grimstad, R. Huber, S. Sataluri, and M. Wahl. Naming Plan for Internet Directory-Enabled Applications. *IETF Request For Comments*, 2377, September 1998.

[Goo00]   G. Good. The LDAP Data Interchange Format (LDIF) - Technical Specification. *IETF Request For Comments*, 2849, June 2000.

[Gre02]   B. Greenblatt. *Building LDAP-Enabled Applications with Microsoft's Active Directory and Novell's NDS*. Prentice Hall PTR, 2002.

[GSB$^+$04]   R. Guida, R. Stahl, T. Bunt, G. Secrest, and J. Moorcones. Deploying and Using Public Key Technology: Lessons Learned in Real Life. *IEEE Security & Privacy*, 2(4):67–71, 2004.

[Gut00]   P. Gutmann. A Reliable, Scalable General-purpose Certificate Store. In *Proceedings of the 16th Annual Computer Security Applications Conference (ACSAC'00)*, pages 278–287, December 2000.

[HFPS99]   R. Housley, W. Ford, W. Polk, and D. Solo. Internet X.509 Public Key Infrastructure Certificate and CRL Profile. *IETF Request For Comments*, 2459, January 1999.

[HM02]   J. Hodges and R. Morgan. Lightweight Directory Access Protocol (v3): Technical Specification. *IETF Request For Comments*, 3377, September 2002.

[HMW00]   J. Hodges, R. Morgan, and M. Wahl. Lightweight Directory Access Protocol (v3): Extension for Transport Layer Security. *IETF Request For Comments*, 2830, May 2000.

[HP01]   R. Housley and T. Polk. *Planning for PKI*. John Wiley & Sons, Inc., 2001.

[HPFS02]   R. Housley, W. Polk, W. Ford, and D. Solo. Internet X.509 Public Key Infrastructure Certificate and Certificate Revocation List (CRL) Profile. *IETF Request For Comments*, 3280, April 2002.

[HS97]   T. Howes and T. Smith. The LDAP URL Format. *IETF Request For Comments*, 2255, December 1997.

[IET]   PKIX Working Group IETF. Public–Key Infrastructure IETF Working Group. http://www.ietf.org/html.charters/pkix-charter.html. Internet Drafts are updated often and on this site the newest status can be obtained.

[IT97]      Recommendation X.509 ITU-T. Information Technology - Open Systems Interconnection - The Directory: Authentication Framework. August 1997.

[KLW04]     V. Karatsiolis, M. Lippert, and A. Wiesmaier. Using LDAP Directories for Management of PKI Processes. In *Proceedings of Public Key Infrastructure: First European PKI Workshop: Research and Applications, EuroPKI 2004*, volume 3093 of *Lecture Notes in Computer Science*, pages 126–134, June 2004.

[KWG+98]    S. Kille, M. Wahl, A. Grimstad, R. Huber, and S. Sataluri. Using Domains in LDAP/X.500 Distinguished Names. *IETF Request For Comments*, 2247, January 1998.

[Leg01a]    The Legislator. Gesetz über Rahmenbedingungen für elektronische Signaturen und zur Änderung weiterer Vorschriften. *Bundesgesetzblatt Jahrgang 2001 Teil I*, Nr. 22:876–884, 21. Mai 2001.

[Leg01b]    The Legislator. Verordnung zur elektronischen Signatur (Signaturverordnung-SigV). *Bundesgesetzblatt Jahrgang 2001 Teil I*, Nr. 59:3074–3084, 21. November 2001.

[MAM+99]    M. Myers, R. Ankney, A. Malpani, S. Galperin, and C. Adams. X.509 Internet Public Key Infrastructure Online Certificate Status Protocol - OCSP. *IETF Request For Comments*, 2560, June 1999.

[MSNP04]    R. Megginson, Ed., M. Smith, O. Natkovich, and J. Parham. Lightweight Directory Access Protocol (LDAP) Client Update Protocol (LCUP). *IETF Request For Comments*, 3928, October 2004.

[Mye97]     J. Myers. Simple Authentication and Security Layer (SASL). *IETF Request For Comments*, 2222, October 1997.

[Ram04]     B. Ramsdell. Secure/Multipurpose Internet Mail Extensions (S/MIME) Version 3.1 Message Specification. *IETF Request For Comments*, 3851, July 2004.

[TT04]      T7 and TeleTrust. Common ISIS–MTT Specifications for Interoperable PKI Applications - Version 1.1. http://www.t7-isis.de/ISIS-MTT/isis-mtt.html, 2004.

[Wah97]     M. Wahl. A Summary of the X.500(96) User Schema for use with LDAPv3. *IETF Request For Comments*, 2256, December 1997.

[WAHM00]    M. Wahl, H. Alvestrand, J. Hodges, and R. Morgan. Authentication Methods for LDAP. *IETF Request For Comments*, 2829, May 2000.

[WCHK97]    M. Wahl, A. Coulbeck, T. Howes, and S. Kille. Lightweight Directory Access Protocol (v3): Attribute Syntax Definitions. *IETF Request For Comments*, 2252, December 1997.

[WHK97]     M. Wahl, T. Howes, and S. Kille. Lightweight Directory Access Protocol (v3). *IETF Request For Comments*, 2251, December 1997.